\documentclass[aps,pra,twocolumn,superscriptaddress]{revtex4-1}
\usepackage{graphicx}
\usepackage{amsmath}
\usepackage{color}
\usepackage{soul}
\usepackage{color, xcolor}
\usepackage{ulem}
\usepackage{dcolumn}
\usepackage{bm}
\usepackage[
   bookmarks=true,
   colorlinks=true,
   hyperindex=true,
   citecolor=blue,
   linkcolor=blue,
   urlcolor=red
]{hyperref}

\begin{document}

\title{Enhanced Atom-by-Atom Assembly of Defect-Free Two-Dimensional Mixed-Species Atomic Arrays}

\author{Ming-Rui Wei}
\affiliation{State Key Laboratory of Magnetic Resonance and Atomic and Molecular Physics, Innovation Academy for
Precision Measurement Science and Technology, Chinese Academy of Sciences, Wuhan 430071, China}
\affiliation{School of Physical Sciences, University of Chinese Academy of Sciences, Beijing 100049, China}

\author{Kun-Peng Wang}
\email{wangkunpeng@wipm.ac.cn}
\affiliation{State Key Laboratory of Magnetic Resonance and Atomic and Molecular Physics, Innovation Academy for
Precision Measurement Science and Technology, Chinese Academy of Sciences, Wuhan 430071, China}

\author{Jia-Yi Hou}
\affiliation{State Key Laboratory of Magnetic Resonance and Atomic and Molecular Physics, Innovation Academy for
Precision Measurement Science and Technology, Chinese Academy of Sciences, Wuhan 430071, China}
\affiliation{School of Physical Sciences, University of Chinese Academy of Sciences, Beijing 100049, China}

\author{Yi Chen}
\affiliation{State Key Laboratory of Magnetic Resonance and Atomic and Molecular Physics, Innovation Academy for
Precision Measurement Science and Technology, Chinese Academy of Sciences, Wuhan 430071, China}
\affiliation{School of Physical Sciences, University of Chinese Academy of Sciences, Beijing 100049, China}
\author{Peng Xu}
\affiliation{State Key Laboratory of Magnetic Resonance and Atomic and Molecular Physics, Innovation Academy for
Precision Measurement Science and Technology, Chinese Academy of Sciences, Wuhan 430071, China}
\affiliation{Wuhan Institute of Quantum Technology, Wuhan 430206, China}
\author{Jun Zhuang}
\affiliation{State Key Laboratory of Magnetic Resonance and Atomic and Molecular Physics, Innovation Academy for Precision Measurement Science and Technology, Chinese Academy of Sciences, Wuhan 430071, China}

\author{Rui-Jun Guo}
\affiliation{School of Information Engineering and Henan Key Laboratory of Laser and Opto-Electric Information Technology, Zhengzhou University, Zhengzhou 450001, China}

\author{Min Liu}
\affiliation{State Key Laboratory of Magnetic Resonance and Atomic and Molecular Physics, Innovation Academy for
Precision Measurement Science and Technology, Chinese Academy of Sciences, Wuhan 430071, China}
\author{Jin Wang}
\affiliation{State Key Laboratory of Magnetic Resonance and Atomic and Molecular Physics, Innovation Academy for
Precision Measurement Science and Technology, Chinese Academy of Sciences, Wuhan 430071, China}
\affiliation{Wuhan Institute of Quantum Technology, Wuhan 430206, China}

\author{Xiao-Dong He}
\email{hexd@apm.ac.cn}
\affiliation{State Key Laboratory of Magnetic Resonance and Atomic and Molecular Physics, Innovation Academy for
Precision Measurement Science and Technology, Chinese Academy of Sciences, Wuhan 430071, China}
\affiliation{Wuhan Institute of Quantum Technology, Wuhan 430206, China}

\author{Ming-Sheng Zhan}
\affiliation{State Key Laboratory of Magnetic Resonance and Atomic and Molecular Physics, Innovation Academy for
Precision Measurement Science and Technology, Chinese Academy of Sciences, Wuhan 430071, China}
\affiliation{Wuhan Institute of Quantum Technology, Wuhan 430206, China}
\date{\today}

\begin{abstract}

Defect-free single atom array in optical tweezers  is a promising platform for scalable quantum computing, quantum simulation, and quantum metrology. Extending single-species array to mixed-species one promise to offer new possibilities. In our recent proof of principle realization of defect-free two-dimensional assembly of mixed-species $^{85}$Rb  ($^{87}$Rb) atom arrays [C. Sheng et al.\href{https://journals.aps.org/prl/abstract/10.1103/PhysRevLett.128.083202}{{\color{blue} Phys. Rev. Lett. 128, 083202(2022)}}],  the filling fractions were limited by the imperfect transfer of atoms and the occurrence of logjams during the atom rearrangement. In order to scale up the  size of defect-free mixed-species atom array,  we  scale up the tweezer array and  improve the atom transfer, and upgrade the heuristic heteronuclear algorithm so as to facilitate multiple rearrangement cycles.  Consequently,  we successfully  create defect-free  atom arrays with 120 mixed-species single atoms.  The corresponding filling fraction and defect-free probability are improved to be  98.6(1)\% and  14(2)\%, respectively.  It is anticipated that the enhanced algorithm can be extended to other combinations of atomic species, and this mixed-species atom array is readily for studies of many-body physics, quantum error correction, and quantum metrology.

\end{abstract}

\maketitle

\section{introduction}

In recent years,  the platform of neutral atoms confined in a closely spaced array of optical tweezers (OTs) ~\cite{Kim2016,Barredo2016,Endres2016,Barredo2018,Kumar2018,Norcia2018,Cooper2018,Jenkins2022,Ma2022} has rapidly progressed to relate to numerous fundamental research topics, ranging from quantum simulations of many-body physics~\cite{9Browaeys2020,WuCPB2021,Kaufman2021,Zoller2023}, quantum computing~\cite{Saffman2019,Bluvstein2022,Graham2022,10Bluvstein2023}, metrology~\cite{Madjarov2019,Covey2019,Norcia2019,Young2020}, ultracold collisional physics~\cite{11XuP2015,Reynolds2020,Weyland2021,Zhuang2024} and and association and quantum control of single molecules~\cite{Liu2018,12HeXD2020,Zhang2022,Cornish2023,Ruttley2024,Picard2024,Cheuk2023,Doyle2023,KKNi2024,Cornish2024}. In particular,  the mixed-species array containing two different neutral atoms has  gradually manifested itself as a compelling platform for quantum simulation and quantum computation~\cite{Beterov2015,Zeng2017,Guo2020,Sheng2022,Singh2022}.  Since, akin to the use of mixed-species trapped ions~\cite{Bruzewicz2019}, in such a dual-species architecture, the atomic resonant transition wavelengths differ substantially from each other so that it allows for the independent control of two different sets of qubits, and readout, leading to promise for quantum error correction~\cite{Saffman2016}. For example, it has been demonstrated that the second species atoms can act as sepctators of magnetic field environment and allow mid-circuit measurements for the qubit species atoms~\cite{Singh2023}. Moreover, it also allows for selective tuning of inter- and intra-species~\cite{Beterov2015,Anand2024} Rydberg interactions, for implementing native multi-qubit gates ~\cite{Farouk2023} and buffer-atom-mediated quantum logic gates~\cite{Sun2024} and  for obtaining more flexible Hamiltonian engineering~\cite{Lan2016,Chepiga2024,SUSY2024,Liu2024}.

Scaling up the system size for the atom-array platforms would greatly benefit for the aforementioned potential applications. For single species array, extending from hundreds of atoms~\cite{Mello2019,Ebadi2021,Scholl2021,Tian2023} to thousands of atoms ~\cite{Schlosser2023,Pause2024,Norcia2024,Gyger2024,Manetsch2024,Pichard2024} is currently the subject of a major research effort. In contrast, it is acknowledged that since the first proof-of-principle demonstration of a defect-free dual-isotope Rb atom array
was realized in a 64-tweezer array~\cite{Sheng2022} until the present, the sizes of defect-free mixed-species atom arrays currently remains small, including the very recent work of isotopes of 4$\times$4 Yb atoms~\cite{Nakamura2024}.

In the context of dual-isotope atom arrays preparation experiments,  the atoms are not selectively trapped by the OTs and randomly loaded into the each site of OT array. As a result, the occurrence of misplaced atoms (for example, a $^{85}$Rb target site is occupied with a $^{87}$Rb atom) will show up, leading to that the sorting-atom algorithm applicable to dual-species atom arrays is much more complicated than single-species atom algorithm. Along this line, by designing a sorting algorithm, named the heuristic heteronuclear algorithm (HHA),  with near-fewest sorting-atom moves, a set of 4$\times$6 defect-free dual-alkali-isotopes  ($^{87}$Rb and $^{85}$Rb) atom arrays has been realized in a 64-tweezer array, and the associated success probability is about 5\% with one rearrangement cycle, which is mainly limited by the atom preservation efficiency and transport efficiency~\cite{Sheng2022}. And in this work, the atoms are constrained to move along the rows and columns spanned by the static tweezers, leading to that different types of atoms will block each other in the moving path. Consequently, the current HHA is facing a limitation that the success rate of sorting path calculation decays exponentially with the increasing number of atoms. In order to bypass this limitation,  a heuristic connectivity optimization algorithm (HCOA) based on undirected graph has been proposed and numerically tested~\cite{Tao2022}. Furthermore, Nakamura et al. have recently instead adopted trajectories that pass between adjacent rows and columns of traps and extended the parallel sort-and-compression algorithm to apply to a dual-isotope Yb atom array. However, the reported success probability of making a 4$\times$4 defect-free array from 10 $\times$10 array is only 1\%, which is interpreted by the strong atom loss during transportation~\cite{Nakamura2024}. We note that the limitation of the mono-use of such a type of moves has also been pointed out in the Ref.\cite{Schymik2020}.

In order to scale up the mixed-species atom array, this study employs a high-power laser in conjunction with a spatial light modulator to increase the scale of the tweezer array first, then mitigate the loss of atomic transportation by upgrading the movable tweezer (MT) experimental apparatus, and enhance the HHA by adding the diagonal trajectories so as to remove the occurrence of logjams.  Experimentally,  we successfully create defect-free atom arrays with 120 dual-isotope single atoms. The corresponding filling fraction and defect-free probability are respectively improved to be  98.6(1)\% and  14(2)\% by running multiple rearrangement cycles. In the end, we find out the defect-free probability is mainly limited by  the residual heating from atom fluorescence imaging.

\section{Upgraded experimental setup and defect-free homonuclear array assembly}
\begin{figure}[htbp]
\centering
\includegraphics[width=9cm]{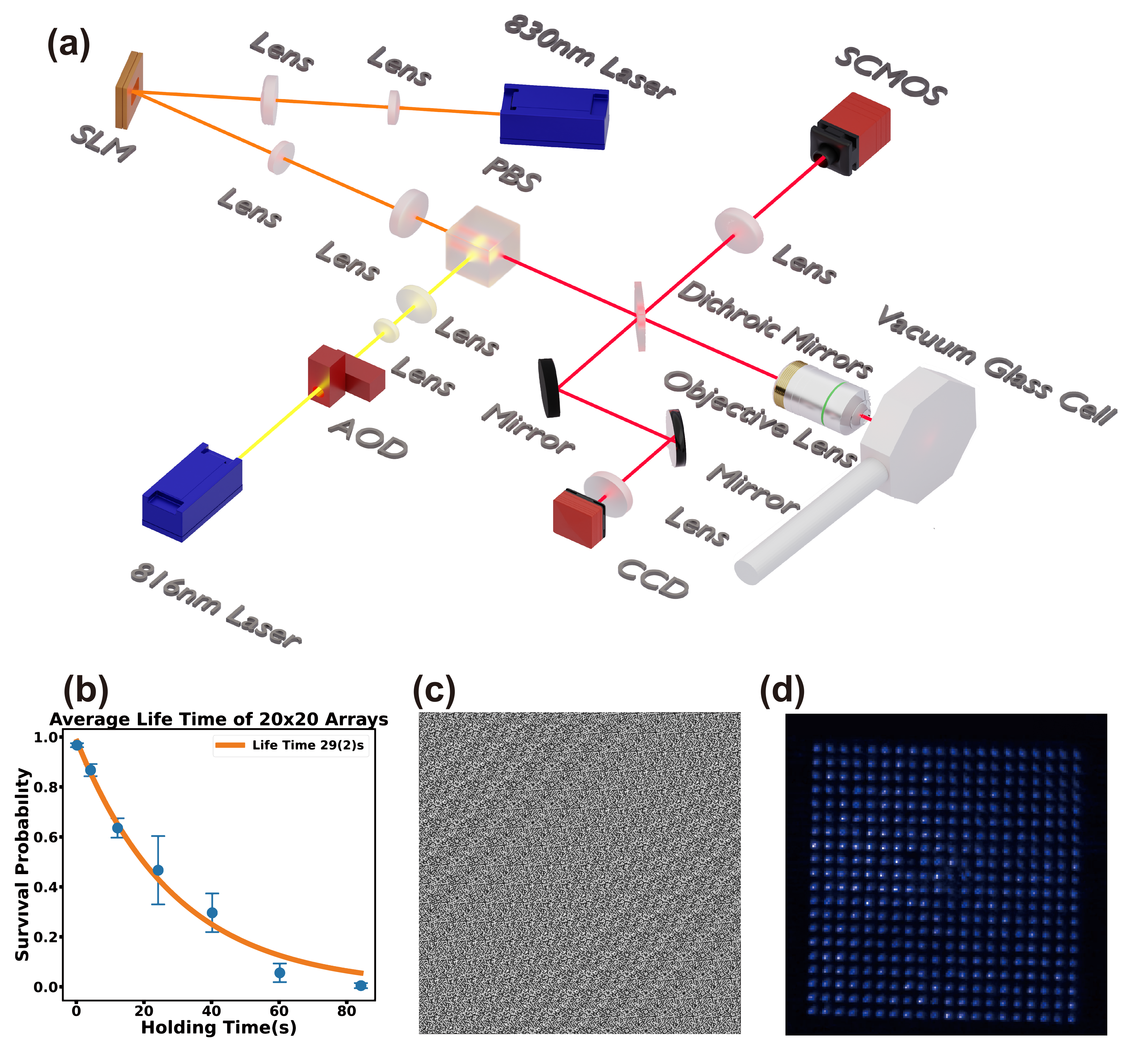}
\caption{Schematic diagram of upgraded experimental setup, tweezer array, phase pattern and life time measurement. (a) Schematic of apparatus. Illustration indicate that spatial light modulator (SLM) and 830 nm-Laser are utilized for creating a 20$\times$20 square optical tweezers array with a 5.4 $\mu m$ spacing, and feedback optimization is performed via the charge-coupled device. A pair of acousto-optic deflector (AOD) with a right angle relative orientation generate movable optical tweezers with 816nm-Laser for rearrangement, strongly focused into the vacuum cavity, and employing scientific complementary metal-oxide-semiconductor (SCMOS) camera on one side to capture atomic fluorescence images. (b) Averaged fluorescence images of the $^{87}$Rb atom trapped in a 400-site tweezer revivor. (c) The corresponding phase pattern on the SLM for making a 400-site tweezer revivor. (d) The measured life time of single atoms in the array.}
\label{fig:fig1}
\end{figure}

Figure 1 (a) schematically shows the upgraded  experimental setup compared with our previous tweezer array setup~\cite{Guo2020,Sheng2022}. The corresponding vacuum pressure level is improved to be below 1E-8 Pa level in a room temperature environment so as to reduce the impact of background gas collisions on the atoms in the array. Consequently, the  atomic trapping lifetime is prolonged to about 29$\pm$2 s in optical tweezers, as shown in Fig.\ref{fig:fig1} (d). This represents nearly a fourfold increase from the 7-second lifetime of atoms in the previous system after applying continuous laser cooling~\cite{sheng2021efficient}.  The tweezer laser beams are focused with a high numerical aperture (NA = 0.6) microscope into a glass cell, which is braodband high transmittance. In order to scale up the optical tweezer array, we incorporate a high power low-noise  laser (830 nm) with a spatial light modulator (SLM). Given an working 1.8 W laser power, we produce a static tweezer array with 20$\times$20 sites by using the weighted Gerchber Saxton (WGS) algorithm to generate the  holograms for SLM~\cite{kim2019large,bianchi2010real,di2007computer,samoylenko2020single}, as shown in Fig.1 (c). The resulting averaged fluorescence images of the $^{87}$Rb atom trapped in a 400-site tweezer revivor is shown in Fig.1 (d). The estimated beam waist for each tweezer is about 0.8 $\mu$m. The typical trapping oscillation frequency is 2$\pi \times$ 100 kHz (2$\pi \times$20 kHz) for the radial (axial) dimension. To improve the utilization rate of the dipole light and reduce the impact of stray light and higher-order stray light from the zeroth-order light on the first-order light we use, we adopted the method of stacking Fresnel lenses to separate the diffracted light from the zeroth-order light. To optimize the uniformity of the resulting tweezer array, a portion of the tweezer laser before entering the strongly focusing system is guided to the CCD for feedback. The resulting uniformity of the optical traps can reach 98\%.

To transport and rearrange atoms into a desired pattern,  a superposed movable tweezer (MT) is build from an additional 816-nm laser, which is deflected with a two-dimensional acoustic-optical deflector (AOD, AA opto-electronic, DTSXY-400-830) . This MT is combined with the tweezer array via  a polarizing beam splitter. Compared with our previous setup, we use a programable arbitratry waveform generator (AWG, Keysight M3202A) to generate rf signals with much lower phase-noise to drive the AOD. The needed rf signals are generated in 14-bit data point segments (1 GS/s) and stored on the RAM of AWG and programmed to output with specific sequences.This AWG allows the preparation of a filled tweezer at target sites via transporting a single atom from another filled tweezer with the MT in typically 1.2 ms. The corresponding control sequence of MT is that: First, the trap depth of MT is adiabatically increasing to 2.3 mK after the MT overlaps a filled reservoir tweezer in $0.2$ ms; then the MT extracts the atom and moves at a definite speed  across the void tweezers to the target site in $0.8$ ms; last the atom in the MT is released into the target site in a adiabatical way in $0.2$ ms. In this process, the associated waveforms of rf signals are optimized in such a way that  the corresponding all the rf frequencies, amplitudes and phases are continuously adjusted to avoid abrupt or discontinuous change, thereby mitigating atomic heating and loss during transfer. To achieve high-fidelity transport and transfer among the array, the position of the mobile tweezer was finely optimized by scanning the rf frequency and maximize the transport efficiency on the first row and and the first column.

In the first set of experiments, we test the validity of the MT experimental apparatus. We run the heuristic cluster algorithm, which has been developed for providing the near-fewest number of moves Nm $\approx$ N (N is the number of defect sites to be filled)~\cite{sheng2021efficient}, to prepare defect-free atom array with arbitrary pattern. An exemplified assembled defect-free atom arrays with $^{87}$Rb are shown in Fig. 2(a). Specifically, for a 10 $\times$10 target structure, the measured probability of defect-free array is about 18.8\% in this system with one rearrangement cycle. This result is better than to the one achieved in the previous one achieved by performing multiple assembly cycles~\cite{Mello2019}, manifesting the benefit of improving the performance of MT.

\begin{figure}[htbp]
\centering
\includegraphics[width=8cm]{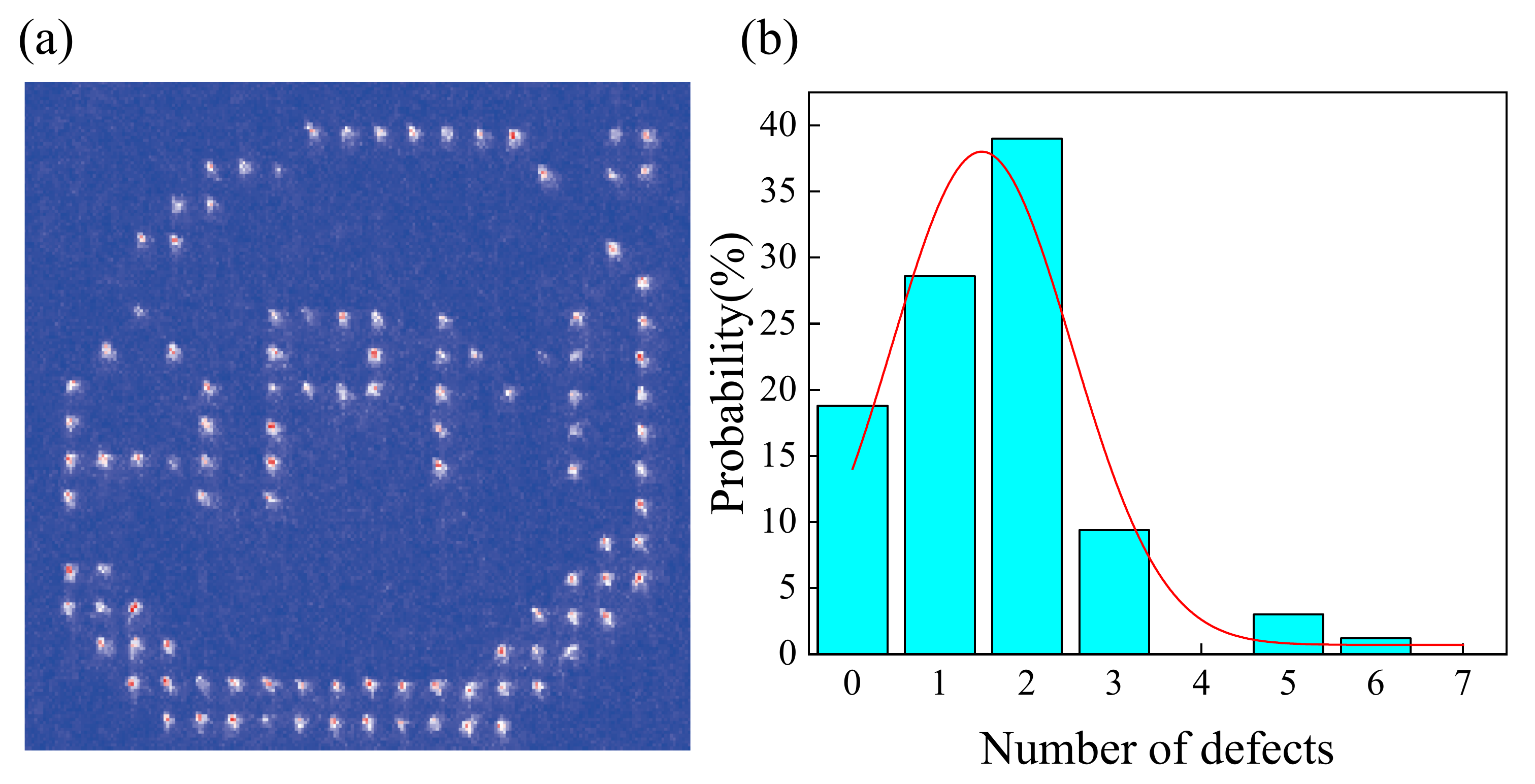}
\caption{Defect-free homonuclear array assembly and the defect-free probability measurement.  (a) Single-shot single-atom resolved fluorescence images of the painting logo of ``APM''. (b) Measured probability distribution of number of defects after one rearrangement cycle. The histogram of data is accumulated by 500 repetitions.  The extracted defect-free probability of defect-free array is about 18.8\%. The shown line is of gaussian fitting, which yields the associated filling fractions of 98.9\%.}
\label{fig:fig2}
\end{figure}

\section{Enhanced heuristic heteronuclear algorithm and defect-free homonuclear array assembly }

\begin{figure}[htbp]
	\centering
	\includegraphics[width=8cm]{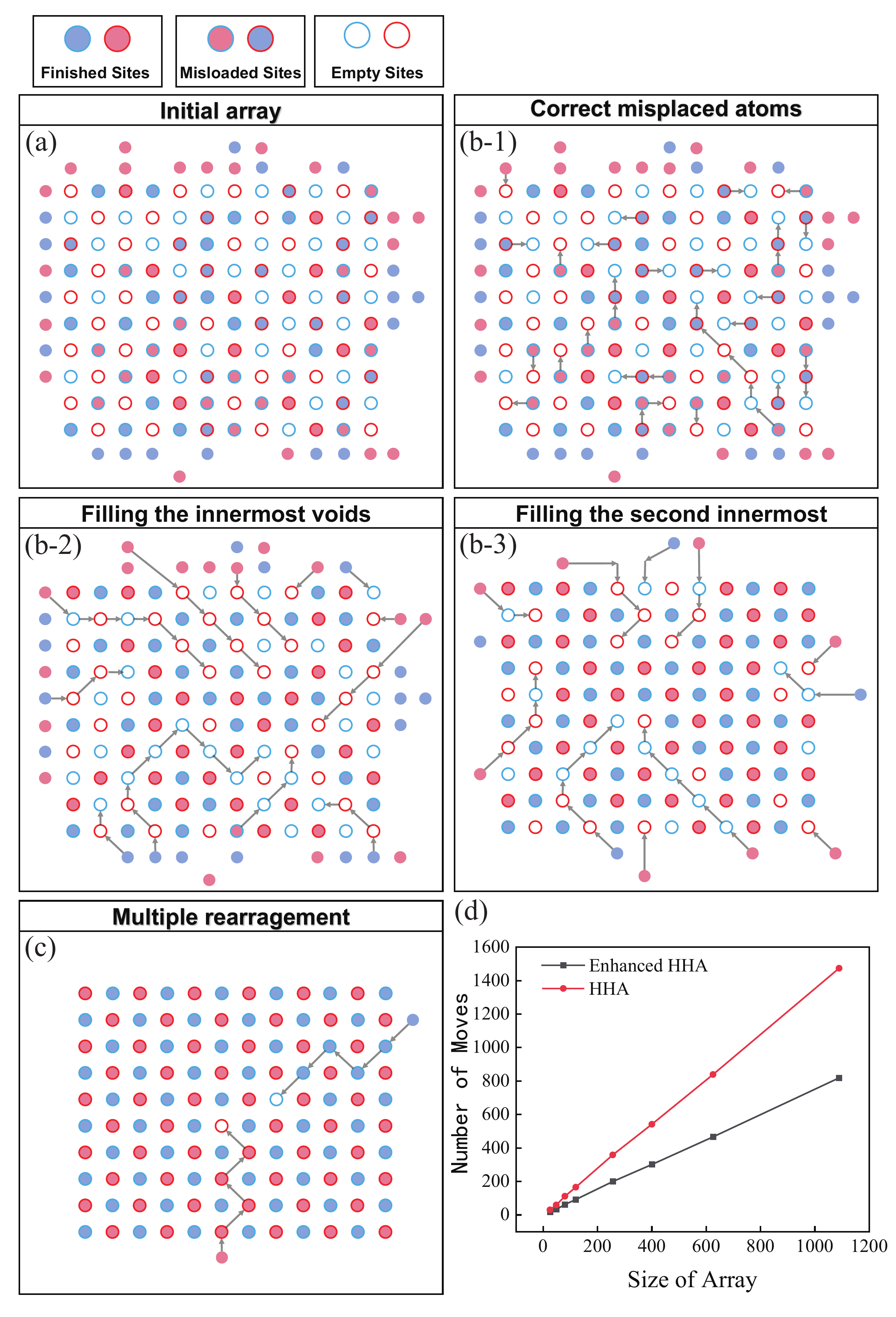}
	\caption{Schematic diagram of the heteronuclear array rearrangement with the enhanced HHA.The blue (red) disks represent single atoms of $^{87}$Rb ($^{85}$Rb) and the blue (red) circles represent the target positions of $^{87}$Rb ($^{85}$Rb). (a) Illustrating a sample initial heteronuclear atomic array after dual-isotope atom loading. (b) Presenting three diagrams rearranged in sequence : (b-1) Correcting the misplaced atoms; (b-2) Filling the innermost voids in different lines; (b-3) Filling the second innermost voids. (c) Illustration of a typical subsequent rearrangement cycle. The deep defects detected are being addressed through the utilization of diagonal movements. (d) The number of moves as a function of the number of filled sites for dual-isotope atom array in zebra configuration. The averaged simulation results are repeated by 100 times with randomly initial configurations for the N $\times$N arrays, which are enlarged by increasing the N. }
	\label{fig:fig3}
\end{figure}

Having demonstrated the benefits of realization of high fidelity transports,  we now present the details of enhanced HHA. The primary enhancement to the HHA involves adding the diagonal moves to the ones along the rows and columns spanned by the static tweezers. It means that the number of working azimuthal angles available for MT movements has been expanded  to eight distinct ones. This improvement allows for effectively addressing situations where atoms were previously unable to move due to being surrounded by other atoms. In our system, the diagonal spacing between any two traps is larger than 7 $\mu$m so that the atom loss due to disturbances of the trap potential can be avoided. Experimentally, such diagonal moves are realized by collaboratively modulating the rf signals of AODX and AODY.

The corresponding flowchart of the enhanced HHA is consistent with the original version, see the Ref. ~\cite{Sheng2022} for details. To illustrate the enhanced HHA, we present a sample sequence of individual rearrangement steps of HHA for making a 120 dual-species checkerboard atom array, which is shown in Fig.\ref{fig:fig3} (a)-(b). In brief,  we first classify the target sites
into finished sites and unfinished sites after the initial loading process. The unfinished sites include misplaced sites and empty sites (voids). First, the innermost voids of each group are filled with the nearest misplaced atoms where the routes connecting them do not have any obstacle atoms, as shown in Fig.\ref{fig:fig3} (b-1) . Then, a repeated process is carried out to fill the new innermost voids, as shown in Fig.\ref{fig:fig3} (b-2) . After several cycles, most of the innermost voids become finished sites. Next, the second (third and so on) innermost voids are filled in the same way , as shown in Fig.\ref{fig:fig3} (b-3) . Compared with the original HHA, the flexible movement of MT significantly reduce the number of moves needed. The numerical simulation indicates that the required number of moves in HHA is reduced by 50\% compared with the old version for making a defect-free 120-atom arrays, as shown in Fig.\ref{fig:fig3}(d).  More importantly, the enhanced HHA enables us to implement subsequent rearrangement cycles to correct for defects arising from the previous cycles, even if these defects are located at the innermost point, as illustrated in Fig.\ref{fig:fig3}(c).  Subsequent cycles necessitate fewer movements and reduced rearrangement time, resulting in diminished loss and augmented filling fractions. This method has been consistently employed for single species atom array assembly~\cite{Endres2016,Kumar2018, Mello2019,Schymik2020,Tian2023}. It is anticipated that, beyond the context of defect-free atom array,  the flexible moves are also essential for the replenishment of a subset of qubits for mid-circuit qubit operations of quantum error correction.

\begin{figure}[htbp]
\centering
\includegraphics[width=9cm]{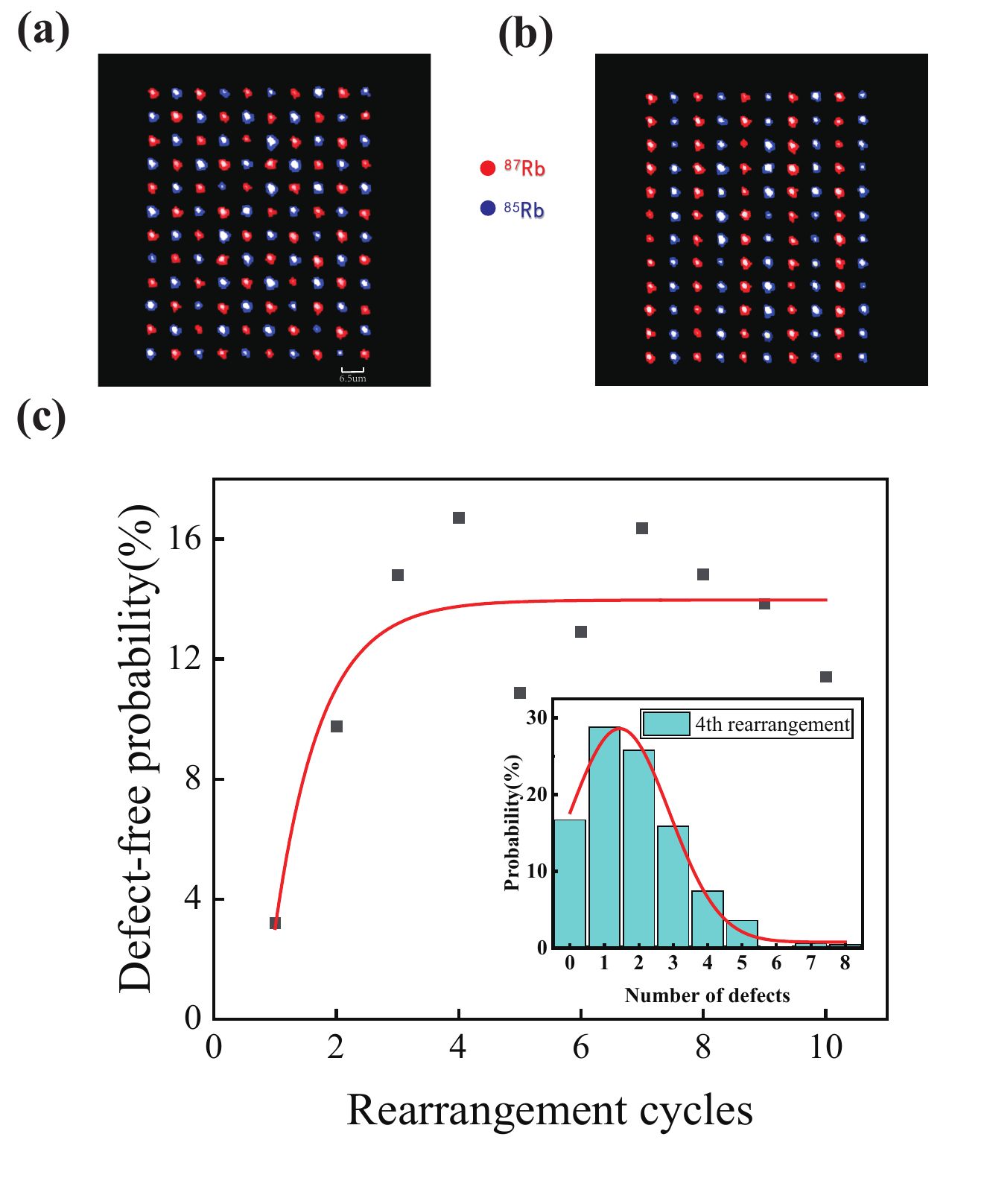}
\caption{The results of the defect-free dual-isotope atom array. (a) Averaged fluorescence image of the dual-isotope checkerboard array with $^{87}$Rb counts in red and $^{85}$Rb counts in blue. (b) Averaged fluorescence image of the dual-isotope zebra array. (c) The dependence of defect-free probability on the number of rearrangement cycles. Shown in the inset is the histogram of data for 4th rearrangements for extracting the defect-free probability and filling fraction, which is accumulated by 500 repetitions.}
\label{fig:fig4}
\end{figure}

Having improved the algorithm, we demonstrate  atom-by-atom assembly of defect-free arbitrary two-dimensional 120-atom dual-species arrays, comprising 60 atoms of each type, which are organized into two distinct patterns: zebra and checkerboard, as illustrated in Fig.~\ref{fig:fig4} (a) and (b), respectively. Specifically, the defect-free probability for the 60-pair defect-free checkerboard array with a single rearrangement cycle is 1.3\%. And the corresponding filling fraction is about 0.9724(9), which is extracted from the similar histogram distribution, as shown in the inset of Fig.~\ref{fig:fig4} (c). We then further improve the filling fractions by running multiple rearrangement cycles. The resulting dependence of defect-free probability on the rearrangement cycles are plotted in  Fig.~\ref{fig:fig4} (c).  It is evident that the defect-free probability experiences a rapid increase in correlation with the number of cycles, subsequently reaching a saturation point of 13.7\%. Following this, subsequent data points exhibit fluctuations. The average of defect-free probability and filling fraction from cycle 4 to 10 are 14(2)\% and 98.6(1)\%,respectively. Obviously, the obtained filling fraction of mixed-species atom array is lower than that of single species one.

\begin{figure}[htbp]
\centering
\includegraphics[width=7cm]{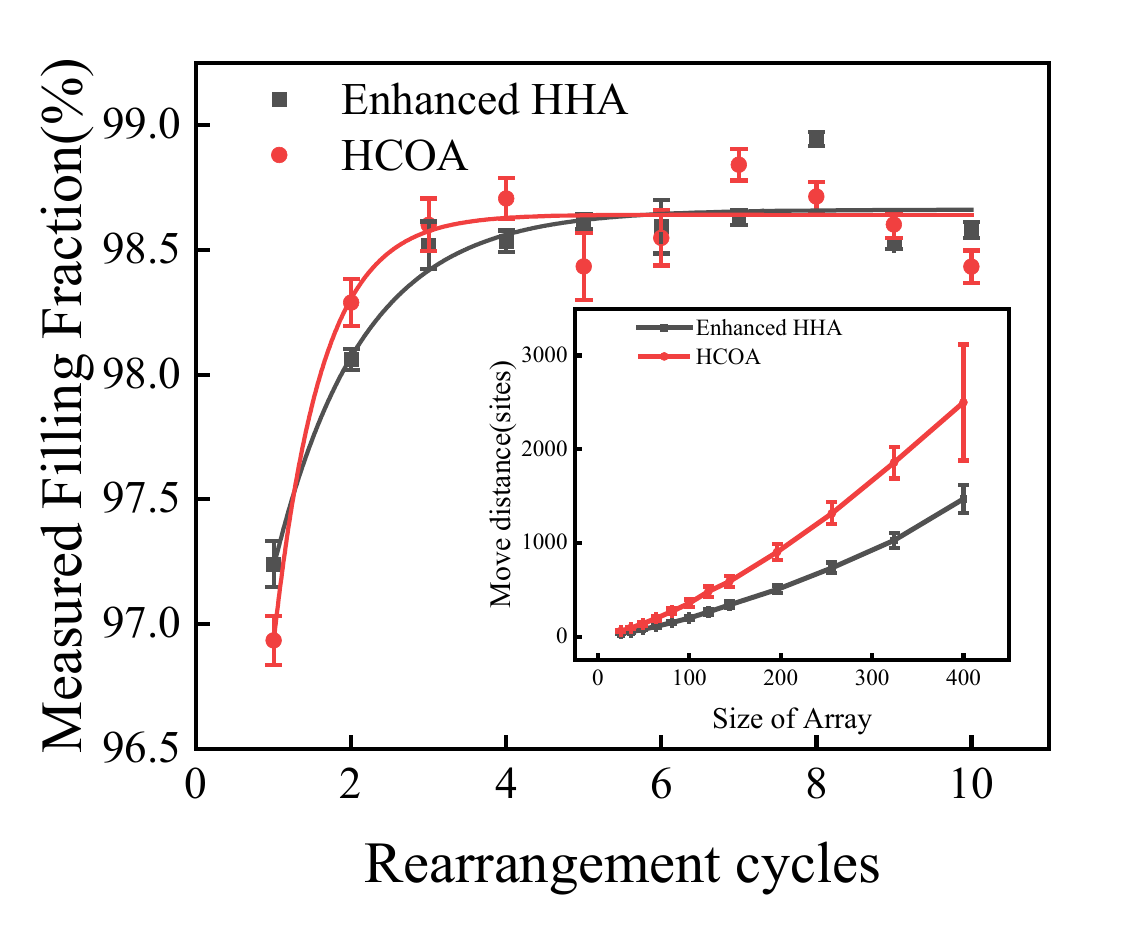}
\caption{The respective dependence of measured filling fraction on the number of rearrangement cycles for the enhanced HHA and HCOA. The accompanying error bars are from fittings. And shown in the inset is the respective dependence of simulated number of sites traversed among the tweezer array for the enhanced HHA and HCOA.}
\label{fig:fig5}
\end{figure}

To understand the associated limitation of filling fraction, we first compared with the aforementioned HCOA~\cite{Tao2022} used for preparing the same heteronuclear arrays. Compared with the enhanced HHA, the HCOA needs pass through more tweezer sites and therefore longer distances, as shown in the inset of Fig.~\ref{fig:fig5}. We find that the resulting dependence of filling fraction on the rearrangement cycles is similar to the case of HHA, as shown in the Fig.~\ref{fig:fig5}. From this data, the extracted average filling fraction is also 98.6(1)\%. Consequently, we infer that the contribution of rearrangement losses is negligible. We then examine the heating from heteronuclear fluorescence imaging, and indeed observe residual cross-talk heating. This is because of that, for heteronuclear fluorescence imaging, we need to successively shine the whole atom array with the probing lights of  $^{87}$Rb and $^{85}$Rb for recording the respective fluorescence, which in turn leads to cross-talk heating between the $^{87}$Rb and $^{85}$Rb. We therefore believe that  the bound on the defect-free array probability is limited by the atomic loss due to the heating from atom fluorescence imaging.
We note that among the dual-species array, the success rate for defect-free $^{87}$Rb($^{85}$Rb) atom sub-array is 26\% (37\%). Compared to the defect-free probability which determined the mean filling fraction of 88.5\% in our previous work~\cite{Sheng2022}, the measured defect-free probability for the 120-atom array presented in this study is approximately  five orders of magnitude higher. With the measured atomic lifetime of 29 s of the array, the whole assembled array can survive by 1 s with a probability of 96\%. Together with the seconds scale coherence times in dual-species array~\cite{Guo2020}, this platform offers a good starting point for quantum information processing.

\section{Conclusion}
In conclusion,  we have scaled up the tweezer array and demonstrated defect-free  arrays of 120 dual-isotope single atoms  with a filling fraction of 98.6(1)\% and defect-free probability of 14(2)\% by running multiple rearrangement cycles with enhanced HHA. The dual-species atom array we have generated is immediately applicable for quantum error correction applications and the exploration of new quantum phases, particularly when integrated with long-range Rydberg atom interactions. High fidelity entanglement for two-qubit and multi-qubit between inter- and intra-species atoms among the array could be achieved with modulated Rydberg pulses~\cite{Fu2022,Sun2024}. In future work aimed at further scaling up the atomic array, several avenues can be explored.  These include mitigating the heating effects caused by imaging, placing the tweezer array within a cryogenic environment~\cite{Schymik2021,Pichard2024}, extending the $\Lambda$-enhanced gray-molasses for dual isotopes~\cite{Brown2019}, and developing sorting algorithms capable of accommodating multiple tweezers in parallel, akin to those designed for single species~\cite{Tian2023,Wang2023}.

\section{acknowledgments}

This work was supported by the National Innovation Program for Quantum Science and Technology of China under Grant No. 2023ZD0300401, the National Natural Science Foundation of China under grants No. 12122412, No. U22A20257, No. 12241410, No. 12121004, No. U20A2074, No. 12261131507, No. 12104414, No. 12104464, the CAS Project for Young Scientists in Basic Research under Grant No. YSBR-055, the National Key Research and Development Program of China under Grant No. 2021YFA1402001, the Major Program (JD) of Hubei Province under Grant No. 2023BAA020, and the Natural Science Foundation of Hubei Province under Grant No. 2021CFA027.



\end{document}